\input harvmac.tex
\input amssym.def
\input amssym.tex

\def\unlockat{\catcode`\@=11}
\def\lockat{\catcode`\@=12}
\unlockat
\def\newsec#1{\global\advance\secno by1\message{(\the\secno. #1)}
\global\subsecno=0\global\subsubsecno=0
\global\deno=0\global\prono=0\global\teno=0\eqnres@t\noindent
{\bf\the\secno. #1} \writetoca{{\secsym}
{#1}}\par\nobreak\medskip\nobreak}
\global\newcount\subsecno \global\subsecno=0
\def\subsec#1{\global\advance\subsecno by1\message{(\secsym\the\subsecno.
#1)}
\ifnum\lastpenalty>9000\else\bigbreak\fi\global\subsubsecno=0
\global\deno=0\global\prono=0\global\teno=0
\noindent{\it\secsym\the\subsecno. #1} \writetoca{\string\quad
{\secsym\the\subsecno.} {#1}}
\par\nobreak\medskip\nobreak}
\global\newcount\subsubsecno \global\subsubsecno=0
\def\subsubsec#1{\global\advance\subsubsecno by1
\message{(\secsym\the\subsecno.\the\subsubsecno. #1)}
\ifnum\lastpenalty>9000\else\bigbreak\fi
\noindent\quad{\secsym\the\subsecno.\the\subsubsecno.}{#1}
\writetoca{\string\qquad{\secsym\the\subsecno.\the\subsubsecno.}{#1}}
\par\nobreak\medskip\nobreak}
\global\newcount\deno \global\deno=0
\def\de#1{\global\advance\deno by1
\message{(\bf Definition\quad\secsym\the\subsecno.\the\deno #1)}
\ifnum\lastpenalty>9000\else\bigbreak\fi
\noindent{\bf Definition\quad\secsym\the\subsecno.\the\deno}{#1}
\writetoca{\string\qquad{\secsym\the\subsecno.\the\deno}{#1}}}
\global\newcount\prono \global\prono=0
\def\pro#1{\global\advance\prono by1
\message{(\bf Proposition\quad\secsym\the\subsecno.\the\prono #1)}
\ifnum\lastpenalty>9000\else\bigbreak\fi
\noindent{\bf Proposition\quad\secsym\the\subsecno.\the\prono}{#1}
\writetoca{\string\qquad{\secsym\the\subsecno.\the\prono}{#1}}}
\global\newcount\teno \global\prono=0
\def\te#1{\global\advance\teno by1
\message{(\bf Theorem\quad\secsym\the\subsecno.\the\teno #1)}
\ifnum\lastpenalty>9000\else\bigbreak\fi
\noindent{\bf Theorem\quad\secsym\the\subsecno.\the\teno}{#1}
\writetoca{\string\qquad{\secsym\the\subsecno.\the\teno}{#1}}}
\def\subsubseclab#1{\DefWarn#1\xdef
#1{\noexpand\hyperref{}{subsubsection}%
{\secsym\the\subsecno.\the\subsubsecno}%
{\secsym\the\subsecno.\the\subsubsecno}}%
\writedef{#1\leftbracket#1}\wrlabeL{#1=#1}}
\lockat

\def\IC{{\Bbb C}}

\def\IR{{\Bbb R}}
\def\IZ{{\Bbb Z}}

\def\pr {\partial}

\def\refb[1] {{(\ref{\#1})}}
\def\eq[1] {{eq.(\ref{\#1}) }}
\def\bear {{\begin{array}}}

\def\qq[1] {{\frac{\overline{\pr ^2 {\cal F}}}{\pr z^{\overline {\#1}}}}}

\def\inv[1] {{{\#1}^{-1}}} 

\def\az {{\overline{z}}}

\def\LL {{\cal L}}

\def\pr {{\partial}}

\def\az {{\bar z}}

\def\t {{\theta}}

\def\a {{\alpha}}
\def\b {{\beta}}
\def\g {{\gamma}}

\def\l {{\lambda}}

\def\CA {{\cal A}}

\def\CC {{\cal C}}
\def\CD {{\cal D}}

\def\CF {{\cal F}}
\def\CG {{\cal G}}
\def\CH {{\cal H}}

\def\CL {{\cal L}}
\def\CM {{\cal M}}

\def\CO {{\cal O}}

\def\CR {{\cal R}}


\def\Fg{{\frak g}}




\def\Tr{{\rm Tr}}

\font\manual=manfnt \def\dbend{\lower3.5pt\hbox{\manual\char127}}

\def\boxit#1{\vbox{\hrule\hbox{\vrule\kern8pt
\vbox{\hbox{\kern8pt}\hbox{\vbox{#1}}\hbox{\kern8pt}}
\kern8pt\vrule}\hrule}}
\def\mathboxit#1{\vbox{\hrule\hbox{\vrule\kern8pt\vbox{\kern8pt
\hbox{$\displaystyle #1$}\kern8pt}\kern8pt\vrule}\hrule}}
\def\frac#1#2{{#1\over#2}}

\lref\mns{G.~Moore, N.~Nekrasov and S.~Shatashvili, {\it Integrating over
Higgs Branches, }
Commun.\ Math.\ Phys.\ {\bf 209} (2000) 97,
[arXiv:hep-th/9712241].}
\lref\GS{A.~Gerasimov and S.~Shatashvili,
{\it Higgs bundles, gauge theories and quantum groups,}
  [arXiv:hep-th/0609024].}
\Title{ \vbox{\baselineskip12pt \hbox{hep-th/yymmnnn}
\hbox{ITEP-TH-07-XX}
\hbox{HMI-07-08}
\hbox{TCD-MATH-07-15}}}
 {\vbox{
\centerline{Two-dimensional  Gauge Theories}
\vskip 0.5cm
\centerline{and}
\vskip 0.5cm
\centerline{Quantum  Integrable Systems}
\medskip
 }}
\medskip
\centerline{\bf Anton A. Gerasimov $^{1,2,3}$ and Samson L.
Shatashvili $^{2,3,4}$}
\vskip 0.3cm
\centerline{\it $^{1}$ Institute for Theoretical and
Experimental Physics, Moscow, 117259, Russia} \centerline{\it
$^{2}$ School of Mathematics, Trinity
College, Dublin 2, Ireland } \centerline{\it $^3$ The Hamilton
Mathematics Institute TCD, Dublin 2, Ireland}
\centerline{\it $^{4}$ IHES, 35 route de Chartres,
Bures-sur-Yvette, FRANCE}
\vskip 0.3cm
In this paper  the relation between 2d topological gauge theories and Bethe
Ansatz equations is reviewed.\foot{These are notes of the talk delivered
by second author at {\it ``From tQFT to tt* and integrability"},
Augusburg, May 2007. } In addition we present some
new results and clarifications.
We hope the relations discussed here are particular examples of
more general relations between quantum topological fields theories in
dimensions $d\leq 4$ and
quantum integrable systems.

\Date{}
\newsec{Introduction}

In \mns\  a  relation between a certain type of two-dimensional
Yang-Mills theory and the Bethe Ansatz equations for the quantum theory of
the Nonlinear Schr\"{o}inger  equation was uncovered. The topological
Yang-Mills-Higgs theory considered in \mns\
captures the hyperk\"{a}hler geometry of the moduli space of
Higgs  bundles introduced in \ref\Hitchin{
N. J.~Hitchin, {\it The self-duality equations on a Riemann surface},
Proc. London Math. Soc. 55 (1987), 59-126.} by Hitchin.
The further step in the understanding of the role of Bethe Ansatz
equations in two-dimensional gauge theories
was made in \GS\ where the wave functions in gauge theory were
identified with  eigenfunctions of the quantum Hamiltonian in the
$N$-particle
sector for the quantum theory of the Nonlinear Schr\"{o}dinger
equation, constructed in the framework of the coordinate Bethe Ansatz.
This implies the equivalence between two seemingly different
quantum field theories. Moreover, in \GS\ it was argued that this
phenomena is not isolated and  was shown how to generalize
the considerations of \mns\ from the Yang-Mills-Higgs theory to  $G/G$
gauged WZW
model with Higgs fields. The corresponding partition function is expressed
in terms of
solutions of  Bethe Ansatz equations  describing a particular limit
of XXZ  spin chains. One expects that  generic
two-dimensional topological quiver gauge theories with additional
matter fields in various representations should reproduce the
Bethe Ansatz equations of other known quantum systems with finite number
of degrees of freedom.

Let us remark that topological  quantum field theories on the manifolds
fibered over circle $S^1$ can be effectively reduced to
Quantum mechanical systems with finite number of degrees of
freedom. Being perturbed by topological observables the
higher-dimensional topological theories lead to
multi-parametric evolution in the effective, one-dimensional,
quantum theories.  One can expect that their
integrability also is a property of three and four dimensional gauge
theories
among which the topological theories calculating Donaldson invariants
look especially interesting.

\newsec{On symplectic and hyperk\"{a}hler  reductions}

To motivate the introduction of the topological Yang-Mills-Higgs
theory in the next Section we start
recalling  the representation of integrals over
symplectic and hyperk\"{a}hler quotients  of a $G$-manifold $X$
in terms of integrals over $X$. This is basically covered by \mns;
here we give integral over $X$, expectedly invariant under
$SO(3)$-rotations of symplectic structures.

Consider a symplectic manifold $X$ of dimension $d=2n$
with a Hamiltonian  action of a compact Lie group $G$.
Define a volume form on $X$ as
a maximal non-zero exterior product $\rm{vol}_X=\omega_X\wedge \cdots
\wedge \omega_X$ of the symplectic two-form $\omega_X$. Given a
Hamiltonian action of Lie group $G$ one has a
corresponding moment map:
\eqn\mom{\mu:\, X\to \Fg^*,\qquad \Fg={\rm Lie}(G),}
and the symplectic quotient over the coadjoint orbit $\CO_u$  of an
element $u\in \Fg^*$ is given by:
\eqn\sympq{
Y=\mu^{-1}(\CO_u)/G=\mu^{-1}(u)/G_u.}
Here $G_u\subset G$ is a stabilizer of $u$. Let $\pi:\,\mu^{-1}(u)\to
Y=\mu^{-1}(u)/G_u$ be the projection on $Y$ and
${\rm{vol}}_{\mu^{-1}(u)}=\rm{vol}_X/\wedge^{\rm{dim}\Fg} d\mu$
be a volume form on $\mu^{-1}(u)$ induced by the volume form
${\rm{vol}}_X$ on $X$.
Then the volume form $\rm{vol}_Y$ on the quotient $Y$ is given by:
\eqn\volq{
\rm{vol}_Y(y)=\frac{1}{\rm{Vol}(G_u)}\pi_*({\rm{vol}}_{\mu^{-1}(u)})=
\frac{1}{\rm{Vol}(G_u)}\int_{\pi^{-1}(y)}{\rm{vol}}_{\mu^{-1}(u)}.}
More generally the integral of any function $f$ over $Y$ can be expressed
as an
integral over $X$ as follows:
\eqn\intfq{
<f>=\int_Y \,f \,{\rm{vol}}_Y=\frac{1}{{\rm{Vol}}(G_u)}\,
\int_{X\times \Fg^*}\,F\,\, e^{i<\varphi,(\mu(x)-u)>}
{\rm{vol}}_X\wedge d\varphi,}
where $<,>$ is a natural pairing between $\Fg^*$ and $\Fg$,
$F$ is  an arbitrary smooth extension of $\pi^*(f)$ to a
function on  $X$ and $d\varphi$ as a flat measure on $\Fg^*$.
The integrand is explicitly invariant with respect to the action of $G_u$
and thus
instead of integration  over $G_u$ one can use ``Faddeev-Popov''
gauge fixing procedure:
\eqn\gfvolume{
<f>=\int_{X\times \Fg^*\times \Pi(\Fg_{\IC})}
\,F\, e^{i<\varphi,(\mu(x)-u)>+i<\chi,\nu(x)>+
\sum_{a,b=1}^{{\rm{dim}}(G_u)}\overline{\CC}^a
i_{V^b}d\nu_a(y)\CC^b}\times}
$$\times\,\,
\rm{vol}_X\wedge d\varphi\wedge d\chi\wedge d\overline{\CC}^a\wedge
d\CC,$$
where $V^b\in Vect_X$, $a=1,\ldots {\rm{dim}}(G_u)$
is a bases of vector fields generating the
action of $\frak{g}_u={\rm Lie}(G_u)$,
$\nu_a(x)$, $a=1,\ldots {\rm{dim}}(G_u)$
is a generic set of the gauge fixing functions, $(\overline{\CC},\CC)$
is a pair of odd $\frak{g}_u$-valued ghosts and $\chi$ is
$\frak{g}_u$-valued even.

Now consider a generalization of representation \gfvolume\ to
the case of hyperk\"{a}hler reduction.  Given a manifold $X$ with  three
compatible symplectic structures $\omega_{\alpha}$, $\alpha=1,2,3$ and
an action of a compact Lie group $G$ which is Hamiltonian with respect
to any of the symplectic structures. We have a family of symplectic
structures with an obvious action of $SO(3)$ group through a
three-dimensional representation.
It will be useful to consider a pair of real and complex symplectic
from $\omega_{\IR}=\omega_3$,  $\omega_{\IC}=\omega_1+i\omega_2$.
The hyperk\"{a}hler reduction of $X$ with respect to $G$ is define as
follows:
\eqn\hypered{
Y=(\mu_{\IC}^{-1}(u_c)\cap \overline{\mu_{\IC}^{-1}(u_c)}\cap
\mu^{-1}_{\IR}(u_r))/G_{(u_c,u_r)},}
where $\mu_{\IC}$ and $\mu_{\IR}$ are moment maps for $\omega_{\IC}$
and $\omega_{\IR}$ and $G_{(u_c,u_r)}\subset G$ is a stabilizer of the
triple
$u_r\in \Fg^*$, $u_c\in \Fg^*\otimes \IC$.
The same space (up to some subtleties with
a compactification of the space of stable orbits)
can be obtained as a holomorphic symplectic manifold
by a holomorphic version  of symplectic reduction:
\eqn\holomred{
Y=\mu^{-1}_{\IC}(u_c)/\widetilde{G}_u^c,}
where $\widetilde{G}_u^c\subset G^c$ is a stabilizer of $u_c\in
\Fg^*\otimes \IC$
with respect to the action of the  complexification $G^c$ of $G$. Note
that
$\widetilde{G}^c_u$ is not necessary a complexification of
$G_{u_c,u_r}$. However in the following we will consider
only the case when $\widetilde{G}^c_u$ is indeed a complexification of
$G_{u_c,u_r}$.

It would be natural to  define the integral of a function $f$ over the
quotient $Y$ by an analog of \intfq :
\eqn\wronganalog{
\int_Y f\sim \frac{1}{{\rm{Vol}}(\widetilde{G}^c_u)}\int_{\pi^{-1}(y)}
\pi^*(f){\rm{vol}}_{\mu_{\IC}^{-1}(u_c)},}
where $\pi: \mu_{\IC}^{-1}(u_c)\to Y$,
$\rm{vol}_X=(\omega_{\IC}\wedge \overline{\omega_{\IC}})\wedge
\cdots \wedge (\omega_{\IC}\wedge \overline{\omega_{\IC}})$
is a volume form on $X$.
However this expression is not well defined because the volume of
$\widetilde{G_u^c}$
is not finite and the fibres of the projection $\pi$ are not compact.

To represent integral over the hyperk\"{a}hler quotient
one should consider an analog of \gfvolume.
It is useful to impose gauge-fixing conditions only on a troublesome
non-compact part of $G_u^c$  so that the compact subgroup
$G_{u_c,u_r}\subset G_u^c$ remains unfixed.
The natural choice of the gauge-fixing conditions is a set
 of real moments $\mu_{\IR}$. Thus we arrive at the following
representation: 
\eqn\gfhvolume{
<f>=\int_Y f=\frac{1}{{\rm{Vol}}(G_{u_c,u_r})}\,\times }
$$\times
\int_{X\times (\Fg^*\otimes \IR^3)}\,F\,
e^{i\sum_{\a=1}^3<\varphi_{\a},(\mu^{\a}(x)-u^{\a})>
+\sum_{a,b=1}^{{\rm{dim}}(G_{u_c,u_r})}\overline{\CC}^a
i_{V^b}d\mu_{\IR}^a(y)\CC^b}
{\rm{vol}}_X\wedge d\varphi_{\a}\wedge d\overline{\CC}\wedge d\CC
$$
This integral allows an explicitly $SU(2)$-invariant reformulation:
\eqn\gfhvolumeinv{
<f>=\frac{1}{\rm{Vol}(G_{u_c,u_r})}\,
\int_{X\times (\Fg^*\otimes \IR^3)\times \Pi(\Fg_{\IC})}
\,F\, e^{i\CF(x,\phi,\CC,\overline{\CC})}
\,\,{\rm{vol}}_X\wedge d\varphi_{\a}\wedge d\overline{\CC}^a\wedge
d\CC^b,}
$$
\CF(x,\phi,\CC,\overline{\CC})=\sum_{\a=1}^3<\varphi_{\a},(\mu^{\a}(x)-u^{\a})>
+\sum_{\a,\b,\g=1}^3\,\sum_{a,b=1}^{{\rm{dim}}(G_{u_c,u_r})}\,
\overline{\CC}^a\, \{\mu^a_{\a},\mu^b_{\b}\}_{\g}\CC^b\,\epsilon^{\a
  \b\g}
$$
where $\mu_{\a}$, $\a=1,2,3$ is a triple of moment maps corresponding
to symplectic forms $\omega_{\a}$,
$\{\,,\,\}_{\a}$ are  Poisson brackets
given by inverse of symplectic forms $\omega_{\a}$ and
$F$ is extension to $X$ of a lift of $f$ to
$\mu_{\IC}^{-1}(u_c)\cap \overline{\mu_{\IC}^{-1}(u_c)}\cap
\mu^{-1}_{\IR}(u_r)$.

In next chapter we consider the example of infinite-dimensional
hyperk\"{a}hler quotient
give by Hitchin equation.

\newsec{Topological Yang-Mills-Higgs theory }

In \mns\ a  two dimensional gauge
theory was proposed such that the space of classical solutions
of the theory on a Riemann surface $\Sigma_h$ is closely related with
the cotangent space to the space of solutions of the dimensionally
reduced (to 2d)  four-dimensional self-dual Yang-Mills   equations studied
by  Hitchin
\Hitchin. Let us given a principal $G$-bundle $P_G$ over a Riemann surface
$\Sigma$ supplied with a complex structure. Then  we have an associated
vector bundle ${\rm ad}_{\Fg}=(P_G\times \Fg)/G$ with
the fiber $\Fg=Lie(G)$ supplied with
a coadjoint action of $G$.  Consider the
pairs $(A,\Phi)$ where  $A$ is a connection on $P_G$ and $\Phi$ is a
one-form taking values in ${\rm ad}_{\Fg}$. Then Hitchin equations are
given by:
\eqn\HEQUAT{
F(A)-\Phi \wedge
\Phi=0,\,\,\,\,\,\,\nabla_A^{(1,0)}\Phi^{(0,1)}=0,\,\,\,\,\,\,
 \nabla_A^{(0,1)}\Phi^{(1,0)}=0.}
The space of the solutions has a natural hyperk\"{a}hler structure and
admits  compatible $U(1)$ action. The correlation functions in the theory
introduced in \mns\ can be described by the integrals of the products
of $U(1)\times G$-equivariant cohomology classes over the moduli space
of solutions of \HEQUAT.   Note that the $U(1)$-equivariance  makes the
path integral well-defined.

The  field content of the theory introduced in \mns\ can be described
as follows. In addition to the  triplet $(A,\psi_A,\varphi_0)$
of the  topological  Yang-Mills  theory one has:
\eqn\FCONT{
(\Phi,\psi_{\Phi}):
\,\,\,\,\,\,\,\,\,\,\,\Phi\in
\CA^1(\Sigma, {\rm ad}_{\Fg}),\,\,\,\,\,\,\,\,\,
\psi_{\Phi}\in \CA^1(\Sigma, {\rm ad}_{\Fg})
}
\eqn\FCONTone{
(\varphi_{\pm},\chi_{\pm}):\,\,\,\,\,\,\,\,\,\,\varphi_{\pm}\in
\CA^0(\Sigma, {\rm ad}_{\Fg}),
\,\,\,\,\,\,\,\,\,\,\,\chi_{\pm}\in \CA^0(\Sigma, {\rm ad}_{\Fg})}
where $\Phi$, $\varphi_{\pm}$ are even  and $\psi_{\Phi}$,
$\chi_{\pm}$ are odd fields. We will use also another notations
  $\varphi_{\pm}=\varphi_1\pm i\varphi_2$.

The theory is described by the following path integral:
 \eqn\YMH{
Z_{YMH}(\Sigma_h)=\frac{1}{{\rm Vol}(\CG_{\Sigma_h})}
\int\, D\varphi_0\,D\varphi_{\pm}\,DA\,D\Phi\,
D\psi_A \,D\psi_{\Phi} D\chi_{\pm}\,\, e^{S(\varphi_0,\varphi_{\pm}
,A,\Phi,\psi_A,\psi_{\Phi},\chi_{\pm})},}
where $S=S_0+S_1$ with:
\eqn\ACTBOS{\eqalign{
S_0(\varphi_0,\varphi_{\pm},A,\Phi,\psi_A, & \psi_{\Phi},\chi_{\pm})=
\frac{1}{2\pi}\int_{\Sigma_h} d^2z\,\Tr(i\varphi_0\,(F(A)-
\Phi\wedge \Phi)-c\Phi\wedge *\Phi)+\cr
+ &
\varphi_+\nabla^{(1,0)}_A\,\Phi^{(0,1)}+\varphi_-\nabla^{(0,1)}_A\,\Phi^{(1,0)}}}
and
\eqn\ACTFERM{\eqalign{
S_1(\varphi_0,\varphi_{\pm},A,\Phi,\psi_A,\psi_{\Phi},\chi_{\pm})=
\frac{1}{2\pi}\int_{\Sigma_h} d^2z\,\Tr(\frac{1}{2}\psi_{A}\wedge
\psi_{A}+
\frac{1}{2}\psi_{\Phi}\wedge \psi_{\Phi}+\cr
+\chi_+[\psi^{(1,0)}_A,\Phi^{(0,1)}]+\chi_-[\psi^{(0,1)}_A,\Phi^{(1,0)}]
+\chi_+\nabla^{(1,0)}_A \psi^{(0,1)}_{\Phi}+\chi_-\nabla^{(0,1)}_A
\psi^{(1,0)}_{\Phi})}}
where the decompositions $\Phi=\Phi^{(1,0)}+\Phi^{(0,1)}$ and
$\psi_{\Phi}=\psi^{(1,0)}_{\Phi}
+\psi^{(0,1)}_{\Phi}$ correspond to the  decomposition of
the space of one-forms $\CA^1(\Sigma_h)
=\CA^{(1,0)}(\Sigma_h)\oplus \CA^{(0,1)}(\Sigma_h)$
defined in terms of  a fixed complex structure on $\Sigma_h$.

 The theory is invariant under the action of  the  following even vector
field:
\eqn\UONEACT{
\CL_v\Phi^{(1,0)}=+ \Phi^{(1,0)}, \,\,\,\,\,
\CL_v\Phi^{(0,1)}=-\Phi^{(1,0)},\,\,\,\,\,
\CL_v\psi^{(1,0)}_{\Phi}=+  \psi^{(1,0)}_{\Phi},}
$$\CL_v\psi^{(0,1)}_{\Phi}=- \psi^{(0,1)}_{\Phi}\,\,\,\,\,
\CL_v\varphi_{\pm}=\mp\varphi_{\pm}, \,\,\,\,\,
\CL_v\chi_{\pm}=\pm \chi_{\pm},\,\,\,\,\,\,\,$$
\eqn\GAUGEYM{
\CL_{\varphi_0}\,A=-\nabla_A\varphi_0,\,\,\,\,
\CL_{\varphi_0}\,\psi_A=-[\varphi_0,\psi_A],
\,\,\,\,\CL_{\varphi_0}\,\Phi=-[\varphi_0,\Phi],\,\,\,
\CL_{\varphi_0}\,\psi_{\Phi}=-[\varphi_0,\psi_{\Phi}],}
$$\CL_{\varphi_0}\,\varphi_0=0,\,\,\,\,\,\,\CL_{\varphi_0}\,\varphi_{\pm}=
-[\varphi_0,\varphi_{\pm}],\,\,\,\,\,\,\,\CL_{\varphi_0}\,\chi_{\pm}=
-[\varphi_0,\chi_{\pm}],   $$
and an odd vector field generated by the BRST operator:
$$
QA=i\psi_A,\,\,\,\, Q\psi_A=-\nabla_A\varphi_0,\,\,\,\,Q\varphi_0=0,$$
\eqn\BRST{
Q\Phi=i\psi_{\Phi},\,\,\,\,Q\psi^{(1,0)}_{\Phi}=-[\varphi_0,\Phi^{(1,0)}]+c\Phi^{(1,0)},
\,\,\,\,\,Q\psi^{(0,1)}_{\Phi}=-[\varphi_0,\Phi^{(0,1)}]-c\Phi^{(0,1)},}
$$Q\chi_{\pm}=i\varphi_{\pm},\,\,\,\,Q\varphi_{\pm}=-[\varphi_0,\chi_{\pm}]\pm
c\chi_{\pm}.$$
We have $Q^2=i\CL_{\varphi_0}+c\CL_v$ and $Q$ can be considered as a
BRST operator on the space of $\CL_{\varphi_0}$ and $\CL_v$-invariant
functionals. The action functional of the topological Yang-Mills-Higgs
theory can be represented as a sum of the action functional of the
topological  pure Yang-Mills theory (written in terms of  fields $\phi_0,
A, \psi_A$) and an additional part which can be
represented as a $Q$-anti-commutator:
\eqn\ACTONQCOM{
S_{YMH}=S_{YM}+[Q,\int_{\Sigma_h}\,d^2z \,\Tr\,(\frac{1}{2}
 \Phi \wedge \psi_{\Phi}+\varphi_+
 \nabla^{(1,0)}_A\Phi^{(0,1)}+\varphi_-
 \nabla^{(0,1)}_A\Phi^{(1,0)})]_+.}
  The theory given by \YMH\ is a  quantum field theory whose
correlation functions are given by the intersections pairings of the
equivariant cohomology classes on the moduli spaces
of Higgs bundles.

To simplify the calculations
it is useful to consider  more general  action given by:
\eqn\ACTTWOPAR{
S_{YMH}=S_{YM}+[Q,\int_{\Sigma_h}\,d^2z \,\, \Tr(\frac{1}{2}
 \Phi\wedge \psi_{\Phi}+}$$+\tau_1\,(\varphi_+
 \nabla^{(1,0)}_A\Phi^{(0,1)}+\varphi_-
\nabla^{(0,1)}_A\Phi^{(1,0)})+\tau_2(\chi_+\varphi_-+\chi_-\varphi_+)\,{\rm
  vol}_{\Sigma_h})].$$
Cohomological localization of the functional integral
takes the simplest form for   $\tau_1=0$, $\tau_2\neq 0$.
Note that it is not obvious that the theory for $\tau_1=0$, $\tau_2\neq 0$
is equivalent to that for $\tau_1\neq 0$, $\tau_2=0$.
However taking into account that the action functionals in these two
 cases differ on the equivaraintly exact form and
for $c\neq 0$ the space of fields is
essentially compact one can expect that the theories are equivalent.

For  $\tau_1=0$ the path integrals over $\Phi$, $\varphi_{\pm}$ and
$\chi_{\pm}$ is quadratic. Thus we have for the partition function
the following formal representation:
\eqn\NONLOCOB{
Z_{YMH}(\Sigma_h)=}
$$=\frac{1}{{\rm Vol}(\CG_{\Sigma_h})}
\int DA\,D\varphi_0\,D\psi_A
\,\,e^{\frac{1}{2\pi}\int_{\Sigma_h}d^2z\,\Tr\,(
i\varphi_0 F(A)+\frac{1}{2}\psi_A\wedge \psi_A)}\,\,
{\rm Sdet}_V(ad_{\varphi_0}+ic),$$
where  the  super-determinant is taken over the super-space:
\eqn\supersp{
V=V_{even}\oplus V_{odd}=\CA^0(\Sigma_h,{\rm ad}\Fg)\oplus
\CA^{(1,0)}(\Sigma_h,{\rm ad}\Fg).}
and should be properly understood using
a regularization compatible with $Q$-symmetry of the path integral
(e.g. $\tau_1\neq 0$).
Thus the Yang-Mills-Higgs theory can be considered as a
 pure Yang-Mills theory deformed by a non-local gauge invariant
observable. Note that for $c \rightarrow \infty$ the action functional
\ACTBOS\ reduces
to that of 2d topological Yang-Mills theory, with corresponding
consequences. At the same
time for $c=0$ it is again 2d topological Yang-Mills action but now for
compelxified gauge group,
thus the theory interpolates for non-zero $c$ between topological
Yang-Mills theory for compact and complex gauge groups, which becomes also
apparent from final answers.

The partition function \YMH\ of the Yang-Mills-Higgs theory on a compact
Riemann surface can be calculated using the standard methods of the
cohomological localization.
As in the case of Yang-Mills theory we consider the deformation of
 the action of the theory:
\eqn\DEFACTNone{
\Delta S_{YMH}=-\sum_{k=1}^{\infty}\,t_k\,\,\int_{\Sigma_h}d^2z \,\,
\Tr\,\varphi_0^{k}\,\,{\rm vol}_{\Sigma_h}.}
where we impose the condition that $t_k\neq 0$ for a finite subset of
indexes $k\in\IZ$.

Path integral with the action \ACTTWOPAR\ at $\tau_1=0$ and $\tau_2=1$ is
easily reduced to the integral over abelian gauge fields. The
  contribution of the additional nonlocal observable in
 \NONLOCOB\ can be calculated  as follows.
 The purely bosonic part of the nonlocal observable
 after reduction to abelian fields can be
easily evaluated using any suitable regularization (i.e. zeta
function regularization) and result is the change of the bosonic
part of the abelian action $\int d^2z(\varphi_0)_iF^i(A)$ by:
\eqn\SHITFS{\eqalign{ \Delta S=&\int_{\Sigma_h}\,d^2z
\,\sum_{i,j=1}^N \log\left(\frac{(\varphi_0)_i-(\varphi_0)_j+ic}{
    (\varphi_0)_i-(\varphi_0)_j-ic}
\right) F(A)^i \cr
& +\frac{1}{2}\int_{\Sigma_h}\,d^2z
\,\sum_{i,j=1}^N \log((\varphi_0)_i-(\varphi_0)_j+ic)R^{(2)}\sqrt{g};}}
where $F(A)^i$ is $i$-th component of the curvature
of the abelian connection $A$ and $R^{(2)}(g)$ is curvature on
$\Sigma_h$ for 2d metric $g$
used to regularize non-local observable.  We will use the notation
$(\varphi_0)_i=\l_i$ in remaining part of the paper. This leads to
the unique $Q$-closed
completion. The completion of the term in \SHITFS\ containing the
curvature of the gauge field  is given by the
two-observable $\CO^{(2)}_f$ corresponding to the descendent of the
following function on the Cartan subalgebra isomorphic to $\IR^N$:
 \eqn\INVF{
f(diag(\l_1,\cdots ,\l_N)=\sum_{k,j=1}^N \int_0^{\l_j-\l_k}\,{\rm
arctg}\, \l/c\,d\l. }
Thus, the abelianized action is defined by two-observable descending
from:
\eqn\yy{ I(\l)=\sum_{j=1}^N (\frac{1}{2}
\l_i^2-2\pi  n_j\l_j)+\sum_{k,j=1}^N \int_0^{\l_j-\l_k}\,{\rm
arctg}\,\, \l/c\,d\l.}
On the other hand the term containing the metric curvature $R$
in \SHITFS\ is $Q$-closed and thus does not need any completion.
It can be considered as an integral of the zero observable:
\eqn\ZEROCURV{
\CO^{(0)}=\sum_{i,j=1}^N \log((\varphi_0)_i-(\varphi_0)_j+ic)}
over $\Sigma_h$ weighted by the half of the metric curvature.
Note that the function $I(\l)$ plays important role in Nonlinear
Schr\"{o}dinger
theory which we explain in next section.

After integrating out the fermionic
partners of abelian connection $A$ the standard localization
procedure leads to following final finite-dimensional integral
representation for the
the partition function \mns: \eqn\SDHsum{
Z_{YMH}(\Sigma_h)=\frac{e^{(1-h)a(c)}}{|W|} \int_{\IR^N} d^N\l\,\,\,
\mu(\l)^{h} \,\sum_{(n_1,\cdots,n_N)\in \IZ^N} e^{2\pi i
\sum_{m=1}^N\l_m n_m}\times }
$$\times  \prod_{k\neq j}(\l_k-\l_j)^{n_k-n_j+1-h}
\,\,\prod_{k,j}(\l_k-\l_j-ic)^{n_k-n_j+1-h}\,e^{-\sum_{k=1}^{\infty}
\,t_k\,\,p_{k}(\l)} ,$$
where
\eqn\MEASURE{
\mu(\l)=\det \|\frac{\pr^2 I(\l)}{\pr \l_i \,\pr \l_j}\|,}
and $p_{k}(\l)$ are $S_N$-invariant  polynomial functions of degree
$k$ on $\IR^N$ and $a(c)$ is a $h$-independent constant
defined by the appropriate choice of the regularization  of the functional
integral.
 One can write the $n_i$-dependent parts of the products in
  \SDHsum\ as the exponent of the sum:
\eqn\SDHsumon{ Z_{YMH}(\Sigma_h)=\frac{e^{(1-h)a(c)}}{|W|}
\int_{\IR^N} d^N\l\,\, \mu(\l)^{h}\, \sum_{(n_1,\cdots,n_N)\in \IZ^N}
e^{2\pi i \sum_j n_j\alpha_j(\lambda)}}
$$\times  \prod_{k\neq j}(\l_k-\l_j)^{1-h}
\,\,\prod_{k,j}(\l_k-\l_j-ic)^{1-h}\,e^{-\sum_{k=1}^{\infty}
\,t_k\,\,p_{k}(\l)} ,$$
with notation:
\eqn\yang{e^{2\pi i \alpha_j(\lambda)}=\CF_j(\l)\equiv
e^{2\pi i \l_j}\prod_{k\neq j}\frac{\l_k-\l_j-ic}{\l_k-\l_j+ic}}
 After taking the sum over $(n_1,\cdots,n_N)\in \IZ^N$  using:
\eqn\correc{\eqalign{\mu(\l) \sum_{(n_1,\cdots,n_N)\in \IZ^N} & e^{2\pi i
\sum_j n_j\alpha_i(\lambda)}=\mu(\l)
\sum_{(m_1,\cdots,m_N)\in \IZ^N}\prod_j\delta(\alpha_j(\l)-m_j)\cr
&
=\sum_{(\l^*_1,\cdots,\l^*_N)\in \CR_N}\prod_j\delta(\l_j-\l_j^*)}}
(see definition of $\CR_N$ below)  and integral over
$(\l_1,\cdots ,\l_N)\in \IR^N$
  we see  that only $\alpha_j(\lambda) \in \IZ$,
or the same - $\CF_j(\l)=1$, contribute to the partition
function which now can be written as:
\eqn\SDHPART{
Z_{YMH}(\Sigma_h)=e^{(1-h)a(c)}\sum_{\l\in \CR_N} \,
D_{\l}^{2-2h}\,e^{-\sum_{k=1}^{\infty}
\,t_k\,\,p_{k}(\l)},}
where:
\eqn\SDHDIM{
D_{\l}=\mu(\l)^{-1/2}\prod_{i<j}(\l_i-\l_j)
(c^2+(\l_i-\l_j)^2)^{1/2},}
and the   $\CR_N$ in \correc\ and \SDHPART\ denotes a  set of the
solutions
of the Bethe Ansatz  equations $\CF_j(\l)=1$:
\eqn\BA{
e^{2\pi i\l_j}\prod_{k\neq
  j}\frac{\l_k-\l_j+ic}{\l_k-\l_j-ic}=1,\qquad\,\,\,\,\,\,\,\,\,\,\,\,
k=1,\cdots
,N,}
for the $N$-particle  sector of the quantum theory of
 Nonlinear Schr\"{o}dinger  equation.
 Note that the sum in \SDHPART\ is taken over the
 classes of the solutions  up to action of the
symmetric group on $\l_i$. This set can be  enumerated by the
 multiplets of the integer numbers $(p_1,\cdots, p_N)\in \IZ^N$ such
 that $p_1\geq  p_2\geq \cdots \geq
 p_N$, $\,p_i\in \IZ$.  Thus, the sum in \SDHPART\
is the sum over the same set of partitions
 as in 2d Yang-Mills theory.

\newsec{$N$-particle wave functions in  Nonlinear Schr\"{o}dinger  theory}

The appearance of a particular form of Bethe Ansatz equations \BA\
 strongly suggests the  relevance of  quantum integrable theories
in the description of  wave-functions in topological Yang-Mills-Higgs
theory. Precisely this form of Bethe Ansatz equations \BA\
 arises in the description of the $N$-particle wave functions
for the quantum Nonlinear Schr\"{o}dinger theory with the coupling
constant $c\neq 0$
\ref\LL{E. H. ~Lieb, W. ~Liniger, {\it Exact analysis of an interacting
Bose gas I. The general solution and the ground state}, Phys. Rev. (2),
 {\bf 130} (1963), 1605.},
\ref\BPF{F. A. ~Berezin, G. P. ~Pohil, V. M. ~Finkelberg, {Vestnik MGU}
{\bf
    1}, 1 (1964) 21.},
\ref\Y{C. N. ~Yang, {\it Some exact results for the many-body problems in
    one dimension with repulsive delta-function interaction},
Phys.Rev.Lett. {\bf 19} (1967), 1312.},
\ref\YY{C. N. ~Yang, C. P. ~Yang, {\it Thermodinamics of a one-dimensional
 system of bosons with repulsive delta-function interaction},
J. Math. Phys. {\bf 10} (1969), 1115.}.
In this  section we recall  the standard facts about the
construction of the these wave-functions
using the coordinate Bethe Ansatz. We also discuss the relation with
the representation theory of the degenerate (double) affine Hecke
algebras and the representation theory of the  Lie groups over complex and
$p$-adic numbers.
 For the application of the quantum inverse
scattering method to
 Nonlinear Schr\"{o}dinger theory see
\ref\KMF{P. P. ~Kulish, S. V. ~Manakov, L. D. ~Faddeev, {\it
Comparison of the exact quantum and quasiclassical  results for
the nonlinear Schr\"{o}dinger equation}, Theor. Math. Phys. {\bf 28}
(1976) 615.},
 \ref\FT{L. D. ~Faddeev, L. A. ~Takhtajan, {\it
Hamiltonian Methods in the Theory of Solitons}, Springer-Verlag,
(1980).}.
One can also recommend \ref\LDFONE{L. D.~Faddeev, {\it
How Algebraic Bethe Ansatz works for integrable model},
``Relativistic gravitation and gravitational radiation'', Proceedings of
Les Houches School of Physics (1995) p. 149,
[arXiv:hep-th/9605187].}
as  a quite readable introduction into the Bethe Ansatz machinery.

The Hamiltonian  of Nonlinear Schr\"{o}dinger theory  with a coupling
constant $c$  is given by:
\eqn\NLSH{
\CH_2=\int dx \,(\frac{1}{2}\frac{\pr \phi^*(x)}{\pr x}\frac{\pr \phi(x)}
{\pr x}+c(\phi^*(x)\phi(x))^2),}
with the following Poisson structure for bosonic fields:
\eqn\Poissonstr{
\{\phi^*(x),\phi(x')\}=\delta(x-x').}
The operator of the number of particles:
\eqn\PN{
\CH_0=\int dx \,\phi^*(x)\phi(x),}
commutes with the Hamiltonian $\CH_2$ and thus one can
 solve the eigenfunction problem
in the sub-sector for a given number of particles $\CH_0=N$.
We will consider  the  both the theory on
infinite interval ($x\in \IR$) and its periodic version $x\in S^1$.
The equation for eigenfunctions in the $N$-particle sector has the
following form:
\eqn\NPARTEIG{
(-\frac{1}{2}\sum_{i=1}^N\frac{\pr^2}{\pr
x_i^2}+c\sum_{1\leq i<j\leq N}\delta(x_i-x_j))\Phi_{\l}(x)=2\pi^2
(\sum_{i=1}^N\l_i^2)\Phi_{\l}(x)\,\,\,\,\,\,\,\,i=1,\cdots ,N.}
This equation is obviously symmetric with respect to the action of
symmetric group $S_N$  on the coordinates $x_i$. Thus the solutions are
classified according to the representations of $S_N$.
 Quantum integrability of the Nonlinear Schr\"{o}dinger
theory implies the existence of the complete set of the commuting
Hamiltonian operators. The corresponding eigenvalues
are given by the symmetric polynomials $p_k(\l)$.

Finite-particle sub-sectors of the Nonlinear Schr\"{o}dinger theory can be
described in terms of the representation theory of a particular kind
of Hecke algebra
\ref\MW{S. ~Murakami, M. ~Wadati {\it
Connection between Yangian symmetry and the quantum inverse scattering
method}, J. Phys. A: Math. Gen. {\bf 29} (1993) 7903.},
\ref\Hi{K. ~Hikami, {\it Notes on the structure of the -function
    interacting gas. Intertwining operator in the degenerate affine
    Hecke algebra}, J. Phys. A: Math. Gen  {\bf 31} (1998) No. 4.},
\ref\HO{G. J. ~Heckman, E. M. ~Opdam, {\it Yang's system of
particles and Hecke algebras}, Annals of Math. {\bf 145} (1997), 139.},
\ref\EOS{E. ~Emsiz, E. M. ~Opdam, J. V. ~Stokman,
{\it Periodic integrable systems with delta-potentials},
[arXiv:math.RT/0503034].}.
 Let $R=\{\a_1,\cdots,\a_{l}\}$ be a root system, $W$ - corresponding Weyl
group and
$P$ - a weight lattice. Degenerate affine Hecke algebra $\CH_{R,c}$
associated to $R$
is defined as an algebra with the basis $S_w$, $w\in W$
 and $\{D_{\l},\l\in P\}$  such that $S_w$ $w\in W$ generate
 subalgebra isomorphic to group algebra $\IC[W]$ and the elements
$D_{\l}$, $\l\in P$  generate the group algebra $\IC[P]$
 of the weight lattice $P$. In addition one has the relations:
\eqn\HA{
S_{s_i}D_{\l}-D_{s_i(\l)}\,S_{s_i}=c\,\frac{2(\l,\a_i)}{(\a_i,\a_i)},
\qquad\qquad \,\,\,\,\,i=1,\cdots
,n.}
Here $s_i$ are the generators of the Weyl algebra corresponding to the
reflection with respect to the simple roots $\a_i$. The center of
$\CH_{R,c}$
  is isomorphic to the algebra of $W$-invariant polynomial functions on
$R\otimes\IC$.
The degenerate  affine Hecke algebras were  introduced  by
Drinfeld \ref\Dr{V. G.~Drinfeld, {\it  Degenerate affine Hecke
algebras and Yangians}, Funct. Anal. Appl. {\bf 20} (1986) 58.} and
independently by Lusztig \ref\Lusztig{G.~Lusztig, {\it Affine Hecke
algebras and their
graded version}, J. AMS {\bf 2}, (1989) 3, 599.}.

Below we consider only the case of ${\frak gl}_N$ root system and thus
we have $W=S_N$. Let us introduce the following differential operators
(Dunkle operators
\ref\Dunkl{T. C.~Dunkl, {\it Differential-Difference operators
associated to reflection groups}, Trans. Amer. Math. Soc. {\bf 311},
 no.1 (1989), 167.}):
\eqn\Dunk{
\CD_i=-i\frac{\pr}{\pr x_i}+i\frac{c}{2}\sum_{j=i+1}^N
(\epsilon(x_i-x_j)+1)s_{ij}.}
Here $\epsilon(x)$ is a sign-function and $s_{ij}\in S_N$ is a
transposition
$(ij)$. These operators together with
the action of the symmetric group \Dunk\  provide a
representation of the degenerate affine algebra $\CH_{N,c}$ for
$\Fg={\frak gl}(N)$:
\eqn\REPDAH{
S_{s_i}\to s_i,\,\,\,\,D_i\to \CD_i,\,\,\,i=1,\cdots ,N.}
The image of the quadratic element of the center
is given by:
\eqn\QUADR{
\frac{1}{2}\sum_{i=1}^N\CD_i^2=
-\frac{1}{2}\sum_{i=1}^N\frac{\pr^2}{\pr x_i^2}+c\sum_{1\leq i<j\leq
N}\delta(x_i-x_j)}
and thus coincides with the restriction of the quantum Hamiltonian
on the $N$-particle sector of Nonlinear Schr\"{o}dinger theory on the
infinite interval.

 We are  interested in  $S_N$-invariant solutions of
 \NPARTEIG. They play the role of  spherical vectors (with respect to the
 subalgebra $\IC[W]\in \CH_{N,c}$) in the representation theory
of degenerate affine Hecke algebra.

The eigenvalue problem \NPARTEIG\ allows the equivalent reformulation
as an eigenvalue problem in the domain $x_1\leq x_2 \leq \cdots \leq x_N$
for the differential operator:
\eqn\NPARTEIGFREE{
(-\frac{1}{2}\sum_{i=1}^N\frac{\pr^2}{\pr x_i^2})\Phi_{\l}(x)=2\pi^2
(\sum_{i=1}^N\l_i^2)\Phi_{\l}(x)\,\,\,\,\,\,\,\,i=1,\cdots ,N,}
 with the boundary conditions:
\eqn\BCN{
(\pr_{x_{i+1}}\Phi_{\l}(x)-\pr_{x_{i}}\Phi_{\l}(x))_{x_{i+1}-x_i=+0}
=4\pi c\Phi_{\l}(x)_{x_{i+1}-x_i=0}.}
The solution is given by:
\eqn\wfunc{
 \Phi^{(0)}_{\l}(x)=\sum_{w\in W}\,\,
\prod_{1\leq i<j\leq
N}\left(\frac{\l_{w(i)}-\l_{w(j)}+ic}{\l_{w(i)}-\l_{w(j)}}
\right)\exp(2\pi i\sum_k\l_{w(k)}x_k),}
 or equivalently:
\eqn\WFUNC{
 \Phi^{(0)}_{\l}(x)=\frac{1}{\Delta_{\Fg}(\l)}\,\sum_{w\in W}
 (-1)^{l(w)}\prod_{1\leq i<j\leq
N}(\l_{w(i)}-\l_{w(j)}+ic)\,\exp(2\pi i\sum_k\l_{w(k)}x_k).}
where $\Delta_{Fg}(\l)=\prod_{1\leq i<j\leq N}(\l_i-\l_j)$.
Note that the wave-function is explicitly symmetric
under the action of symmetric group $S_N$ on $\l=(\l_1,\cdots,\l_N)$.
This solution can be also  constructed using the representation
theory of degenerate affine Hecke algebra $\CH_{N,c}$ (see \MW, \Hi, \HO,
\EOS).

Given a solution of the equations \NPARTEIGFREE\ with boundary
conditions \BCN\  $S_N$-symmetric solutions of \NPARTEIG\ on $\IR^N$
  can be represented in the following form:
 \eqn\wfuncone{
 \Phi^{(0)}_{\l}(x)=\sum_{w\in W}
\left(\prod_{i<j}\left(\frac{\l_{w(i)}-\l_{w(j)}+ic\epsilon
(x_i-x_j)}{\l_{w(i)}-
\l_{w(j)}}
\right)\exp(2\pi i\sum_k\l_{w(k)}x_k)\right),}
where $\epsilon(x)$ is a sign-function.
This gives the full set of  solutions of  \NPARTEIG\
for $(\l_1\leq \cdots \leq \l_N)\in \IR^N$ satisfying the
orthogonality condition with respect to the natural pairing:
\eqn\PAIR{
<\Phi_{\l},\Phi_{\mu}>=\frac{1}{N!}\int dx_1\cdots dx_N\,
\overline{\Phi}_{\l}(x)
\Phi_{\mu}(x)=G(\l)\prod_{i=1}^N\delta(\l_i-\mu_i),}
where:
\eqn\NORMFAC{
G(\l)=\prod_{1\leq i<j\leq N}\frac{(\l_i-\l_j)^2+c^2}{(\l_i-\l_j)^2},}
Therefore the  normalized wave functions are given by:
\eqn\NORMWFUNC{
\Phi_{\l}(x)=\sum_{w\in W}(-1)^{l(w)}\,
\prod_{i<j}\left(\frac{\l_{w(i)}-\l_{w(j)}+ic\epsilon
(x_i-x_j)}{\l_{w(i)}-\l_{w(j)}-ic\epsilon(x_i-x_j)}\right)^{\frac{1}{2}}
\exp(2\pi i\sum_k\l_{w(k)}x_k).}
The eigenvalue problem for periodic $N$-particle
Hamiltonian of  Nonlinear Schr\"{o}dinger theory
 can be reformulated in the following way.
Consider the eigenfunction problem for the differential operator:
\eqn\NPARTEIGPeriod{
(-\frac{1}{2}\sum_{i=1}^N\frac{\pr^2}{\pr
x_i^2}+c\sum_{n\in \IZ}\,\sum_{1\leq i<j\leq
N}\delta(x_i-x_j+n))\Phi_{\l}(x)=2\pi^2
(\sum_{i=1}^N\l_i^2)\Phi_{\l}(x)\,\,\,\,\,\,\,\,i=1,\cdots ,N.}
The wave function of the periodic Nonlinear Schr\"{o}dinger equation
are the eigenfunction of \NPARTEIGPeriod\
satisfying the following invariance conditions:
\eqn\INVARAFFINE{
\Phi_{\l}(x_1,\cdots, x_j+1,\cdots, x_N)=\Phi_{\l}(x_1,\cdots, x_N),
\,\,\,\,\, j=1,\cdots ,N,}
$$
\Phi_{\l}(x_{w(1)},\cdots,
x_{w(N)})=\Phi_{\l}(x_1,\cdots,x_N),\,\,\,\,w\in S_N.$$
These are  the conditions of invariance under the action of the affine
Weyl group on the space of wave functions.

The solutions can be obtained imposing the
additional periodicity conditions on the
wave functions \NORMWFUNC. This  leads to the
following  set of the  Bethe Ansatz equations for
$(\lambda_1,\cdots ,\lambda_N)$:
\eqn\BAONE{
\CF_j(\l)\equiv e^{2\pi i \l_j}\prod_{k\neq
j}\frac{\l_k-\l_j-ic}{\l_k-\l_j+ic}=1,\,\,\,\,\,\,\,\,\,
\,\,j=1,\cdots ,N.}
The set of  solutions of these equations can be enumerated by
sets of integer numbers $(p_1 \geq \cdots  \geq p_N)$
- for each ordered set of these integers
there is exactly one solution to Bethe Ansatz equations \YY.
Let us remark that there is the following equivalent
representation for the periodic wave-functions:
\eqn\NORMWFUNCAFF{
\widetilde{\Phi}_{\l}(x)=\sum_{w\in W}(-1)^{l(w)}\,
\prod_{i<j}\left(\frac{\l_{w(i)}-\l_{w(j)}+ic}
{\l_{w(i)}-\l_{w(j)}-ic}\right)^{\frac{1}{2}+[x_i-x_j]}
\exp(2\pi i\sum_k\l_{w(k)}x_k),}
where $[x]$ is an integer part of $x$ defined by the conditions:
$[x]=0$ for $0\leq x<1$ and $[x+n]=[x]+n$. It easy to see that these
wave functions are periodic and descend to the wave functions \NORMWFUNC\
 if $\l=(\lambda_1,\cdots ,\lambda_N)$ satisfy \BAONE.
The  normalized wave functions in the periodic case are given by:
\eqn\PERIODNORM{
\Phi_{\l}^{norm}(x)=\left(\det\|\frac{\pr \log \CF_j(\l)}{\pr
    \l_k}\|\right)^{-1/2}\Phi_{\l}(x)=\mu(\l)^{-1/2}\Phi_{\l}(x).}
Note that the normalization factor is closely  related to the
factor \MEASURE\ arising in the representation of the partition
function of Yang-Mills-Higgs theory. Indeed the function $I(\l)$
introduced in \MEASURE\ is known in the theory of Nonlinear
Schr\"{o}dinger equations as Yang function \YY;
critical points of Yang function are in one to one correspondence with the
solutions
of Bethe Ansatz equations:
\eqn\YangF{ \alpha_j(\l)=\log
\CF_j(\l)=\frac{\pr I(\l)}{\pr \l_j}=n_j.}
 Below we will see that
this is not accidental. Finally note that the  periodic Nonlinear
Schr\"{o}dinger theory has an interpretation in terms of the
representation theory of the degenerate double affine Hecke
algebras introduced by Cherednik \ref\ICH{I. ~Cherednik, {\it A
unification of Knizhnik-Zamolodchikov and Dunkle operators via
affine Hecke algebras}, Invent. Math. J. {\bf 106} (1991), 411.}.
For the details in this regard see \EOS.

\newsec{Wave-function in topological Yang-Mills-Higgs theory}

In this section we provide the evidences for the identification of
a natural bases of wave functions of topological Yang-Mills-Higgs
theory for $G=U(N)$  (given by a
 path integral on a disk with the insertion of observables in the
center)  with the eigenfunctions of the  $N$-particle
Hamiltonian operator  of Nonlinear Schr\"{o}dinger theory.
  First by counting the observables of
the theory we show that the phase space of the Yang-Mills-Higgs
theory can be considered as a deformation of the phase space of
Yang-Mills theory. We compute the cylinder path
integral (Green function) and torus partition function in Nonlinear
Schr\"{o}dinger theory (with all higher Hamiltonians) and show that
latter coincides with torus partition function in Yang-Mills-Higgs
theory (with arbitrary observables turned on). Then, using the explicit
representation
of the partition function on the two-dimensional torus we derive the
transformation properties of the wave functions under large gauge
transformations. They are in agreement with the known explicit
transformation properties of  wave function in Nonlinear
Schr\"{o}dinger theory. Finally, we demonstrate that
eigenfunctions of the  $N$-particle Hamiltonian operator
  of Nonlinear Schr\"{o}dinger theory indeed coincide with
the bases of Yang-Mills-Higgs wave-functions in appropriate
polarization.

\subsec{Local $Q$-cohomology}

We start with a description of the Hilbert space of the
Yang-Mills-Higgs theory using the operator-state correspondence.
In the simplest form the operator-state correspondence is as
  follows. Each operator, by acting on the  vacuum state,
 defines a sate in the Hilbert space. In turn for each state
there is an operator, creating the state from the vacuum state. Moreover,
for the maximal commutative subalgebra of the operators this
correspondence should be  one to one.
For example, the space of local gauge-invariant $Q$-cohomology classes
in topological $U(N)$ Yang-Mills theory is spent, linearly, by the
operators:
 \eqn\OPYM{
\CO^{(0)}_k=\frac{1}{(2\pi i)^k}\Tr\,\varphi^k,}
and  thus this space coincides with the space of  ${\rm Ad}_G$-invariant
regular functions on the Lie algebra ${\frak u}_N$.
 This is in accordance with the description of
the Hilbert space of the theory  given in Section 2.

We would like to apply the same reasoning to  the topological
Yang-Mills-Higgs theory. To get   economical description of the Hilbert
space of
the theory one should find  a maximal (Poisson) commutative
subalgebra of  local $Q$-cohomology classes
(where $Q$ given by \BRST\ acts on the space of functions
invariant under the symmetries generated by \UONEACT, \GAUGEYM).
Obviously  operators \OPYM\ provide non-trivial cohomology
classes. One can show that these operators provide a maximal
 commutative subalgebra for $c\neq 0$ and therefore
 the reduced phase space in Yang-Mill-Higgs system
 can be identified  with a  phase space of  pure Yang-Mills theory.
Thus the  Hilbert space of Yang-Mills-Higgs theory ($c\neq 0$)
can be naturally identified with the Hilbert space of
Yang-Mills theory (identified with $c\to \infty$).
The fact that the Hilbert space of Yang-Mills-Higgs
 theory is the same for all  $c\neq 0$ implies that the bases of
wave-functions for $c\neq 0$ should be
a deformation of the bases for  $c=\infty$.

One should  stress that this reasoning is not applicable to the case $c=0$
The local cohomology for $c=0$ contains additional operators.
 For example, the following operators provide non-trivial cohomology
classes for arbitrary $t\in \IC$ and $c=0$:
\eqn\OPYNC{
\CO^{(0)}_k(t)=\frac{1}{(2\pi i)^k}\Tr(
\varphi_0+t\varphi_+-\frac{t}{2}\chi_+^2)^k.}
This is a manifestation of the fact that $c=0$ theory is a Yang-Mills
  theory for the complexified group $G^c$ and thus its phase space is
  given by $\CM_{\IC}=T^*H^c/W$.

\subsec{Gauge transformations  of wave function}

The discreteness of the  spectrum of  $N$-particle Hamiltonian
operator in periodic Nonlinear Schr\"{o}dinger theory
arises due to periodicity condition on wave functions.
Thus the eigenfunctions in the periodic  case are given by a subset of
eigenfunctions  on $\IR^N$ descending  to $S_N$-invariant
functions on $(S^1)^N$.
The eigenfunction \NORMWFUNC\  of the Hamiltonian operator on
$\IR^N$ represented as sum over elements of symmetric group
of simple wave  functions. For generic eigenvalues each  term of
the sum is multiplied by some function under the shift $x_i\to
x_i+n_i$, $n_i\in \IZ$ of the coordinates. Below we will show how these
multiplicative factors arising in Nonlinear Schr\"{o}dinger theory
can be derived in Yang-Mills-Higgs gauge theory.

We start with a simple  case of  Yang-Mills theory.
The partition function of $U(N)$ Yang-Mills theory on the torus
$\Sigma_1$ is given by:
\eqn\YMMMsumone{
Z_{YM}(\Sigma_1)=\int_{\IR^N/S_N} d^N\l\,\sum_{(n_1,\cdots ,n_N)\in \IZ^N}
e^{2\pi  i \sum_{m=1}^N\l_m n_m}
\,\,e^{-\sum_{k=1}^{\infty}\,t_k\,\,p_{k}(\l)}=}
$$=\sum_{(m_1,\cdots ,m_N)\in P_{++}}
\,\,e^{-\sum_{k=1}^{\infty}\,t_k\,\,p_{k}(m+\rho)}$$
where $P_{++}$ is a set of the dominant weights of $U(N)$.
The sum over $(n_1,\cdots ,n_N)\in \IZ^N$ has a meaning
of the sum over topological classes of
$U(1)^N$-principle bundles on the torus $\Sigma_1$.
 It results in the replacement  of the
integration over $\l$ by a sum over a discrete subset.
This should be compared with the  partition function of dimensionally
 reduced $U(N)$  Yang-Mills theory on $S^1$:
\eqn\YMREDPARTFUNC{
Z_{QM}(S^1)=\int_{\IR^N/S_N} d^N\l
\,\,e^{-\sum_{k=1}^{\infty}\,t_k\,\,p_{k}(\l)}.}
Contrary to the two-dimensional Yang-Mills theory
in the last case we do not have any additional restriction on the
spectrum $(\l_1,\cdots ,\l_N)\in \IR^N/S_N$.
The appearance of the additional sum
can be traced back to the difference between
the Hilbert spaces of dimensionally reduced and non-reduced theories.

 The mechanism of the spectrum restriction via the sum over the
topological sectors
 can be explained in terms of the structure of the Hilbert space of the
theory as follows.  In the Hamiltonian formalism the partition function
on a torus is given by the  trace
of the evolution operator over the Hilbert space of the theory.
Let us consider first the dimensionally reduced $U(N)$ Yang-Mills theory.
The  phase space of the theory is  $T^*\IR^N/S_N$ where we divide over
Weyl group $W=S_N$. To construct the Hilbert
space we quantize the phase space using  the following   polarization.
Consider Lagrangian projection  $\pi: T^*\IR^N\to\IR^N$
supplied with a section. We chose the  coordinates
on the base  as position variables and
the  coordinates on the fibers  as the corresponding  momenta.
Thus  the Hilbert space in this polarization is realized as a space of
$S_N$ (skew)-invariant functions on the base $\IR^N$ of the projection.

Now  consider  two-dimensional Yang-Mills
theory. For the phase space we have  $T^*H/S_N$ where
$H$ is Cartan subgroup. We use $\,$ similar
$\,$ polarization $\,$ associated $\,$ with $\,$
the projection $\,\,\,\,\,$ $\pi: T^*H\to H$.
 Thus the wave functions are $S_N$ invariant functions on
a  torus $H$ or equivalently  the functions on $\IR^N$
invariant under action of the semidirect product of the lattice
$P_0=\pi_1(H)$ and Weyl $W=S_N$ group (i.e. under the action of the
affine Weyl group $W^{aff}$).
The lattice $P_0$ can be interpreted as a lattice of the
 $\IR^N$-valued constant connections on $S^1$ which are  gauge equivalent
to the
zero connection. The corresponding gauge transformations
act on the  wave functions   by the shifts
$x_j\to x_j+ n_j$, $n_j \in \IZ$  of the argument
of the wave functions in the chosen polarization and the wave
functions in two-dimensional Yang-Mills theory can be obtained by the
averaging over this gauge transformations and global gauge
transformations by the nontrivial elements of the normalizer of Cartan
torus $W=N(H)/H$.

It  is possible to relate the averaging
over the topologically non-trivial transformations with the
sum over topological classes of $H$ bundle on the tours.
Note that the maps of $S^1$ to the gauge group $H$
are topologically classified by $\pi_1(H)=\IZ^N$.
Consider  a connection  $A=(A_1,\cdots, A_N)$
 on a $H$ bundle  over  a cylinder  $L$,
$\pr L=S^1_+\cup S^1_-$  such that the holonomies along the boundaries
$S^1_+$ and $S^1_-$ are in the different topological classes
$[(m_1,\cdots,m_N)]\in \pi_1(H)$ and $[(m_1+n_1,\cdots, m_N+n_N)]\in
\pi_1(H)$.
Gluing  boundaries  of the cylinder $L$ we obtain  a torus supplied
with a  connection $\nabla_A$ such that the  first Chern classes of
the bundles corresponding to each $U(1)$-factor  are given by
 $c_1(\nabla_{A_i})=\frac{1}{2\pi i}\int_LF(A_j)=n_j$, $j=1,\cdots, N$.
Thus we see that the sum over the topologically non-trivial gauge
transformations on $S^1$ can be translated into the sum over
topological classes of the $H$-bundles on the torus.

Let us rederive  the partition function of Yang-Mills theory on the
torus \YMMMsumone\ using the averaging procedure. We start with the
dimensionally reduced theory. Let us chose a bases in the Hilbert
space of the dimensionally reduced Yang-Mills theory given by the
$S_N$ skew-invariant eigenfunctions of the quadratic operator
$H^{(0)}_2=\tr \varphi^2$. In the polarization discussed above we
have:
 \eqn\HAMONE{
H^{(0)}_2\,\psi_{\l}(\t)=-\frac{1}{2}\left(\sum_{j=1}^N\frac{\pr^2}{\pr
  \t_j^2}\right)\,\psi_{\l}(\t)=2\pi^2\sum_{j=1}^N\l_i^2\,\psi_{\l}(\t),}
where $(\t_1,\cdots,\t_N)\in \IR^N$.
The set of  normalized skew-invariant eigenfunctions is given by:
\eqn\NORMEIGONEDIM{
\psi_{\l}(\t)=\sum_{w\in S_N}\,(-1)^{l(w)}\,
\exp(2\pi i \sum_{j=1}^N\l_{w(j)} \t_j),\qquad \,\,\,\,\,\,\,\,\,\,\,
(\l_1,\cdots ,\l_N)\in \IR^N/W.}
\eqn\NORMILAIZATION{
\frac{1}{N!}\int_{\IR^N}\,\,d^N\t\,\,
\overline{\psi_{\l}}(\t)\,\psi_{\l'}(\t)=(2\pi)^N
 \sum_{w\in S_N}\,(-1)^{l(w)}\prod_{j=1}^N\delta(\l_{w(j)}-\l'_j)
=\delta^{(S_N)}(\l-\l').}
The integral kernel of the identity operator acting
on the skew-symmetric functions can be represented
(due to translation invaraince
it is the function of difference $\t-\t'$) as:
\eqn\KERNEL{
K_0(\t,\t')=K_0(\t-\t')=\delta^{(S_N)}(\t-\t')=\int_{\IR^N/S_N}\,
d^N\l\,\,
 \overline{\psi_{\l}}(\t)\,\psi_{\l}(\t').}
The partition function of the dimensionally reduced Yang-Mills theory
on $S^1$ is given by the trace of a evolution operator and
can be written explicitly as:
\eqn\TRACEOP{
Z_{QM}(S^1)=Tr\, e^{-t_2 H_2(\hat{p},\hat{q})}=
\int_{(\IR^N\times \IR^N)/S_N}\,d^N \t\, d^N\l\,\,\,\,
 \overline{\psi_{\l}}(\t)\,e^{-t_2H^{(0)}_2(i\pr_\t,\t)}
\,\psi_{\l}(\t)=}
$$
=\int_{(\IR^N\times \IR^N)/S_N}\,d^N \t\, d^N\l\,\,\,\,
 e^{-t_2 p_2(\l)}.$$
The Green function of the theory is:
\eqn\gzero{G_0(\t,\t')=\int_{\IR^N/S_N} d^N\l\,\,\,\,
 \overline{\psi_{\l}}(\t')\,e^{-t_2H^{(0)}_2(i\pr_\t,\t)}
\,\psi_{\l}(\t)=}
$$
=\int_{\IR^N/S_N} d^N\l\,\,\,\, \psi_{\lambda}(\t')
 e^{-t_2 p_2(\l)}\psi_{\lambda}(\t).$$
Up to the infinite factor given by the integral over
$\t=(\t_1,\cdots,\t_N)\in \IR^N$ the integral in \TRACEOP\ coincides with
the
expression \YMREDPARTFUNC\ for the partition function for $t_{i\neq 2}=0$.

Now consider two-dimensional Yang-Mills theory.
In this case we have the periodic eigenvalue problem for \HAMONE. Then for
the normalized eigenfunctions of $H_2$ we have:
\eqn\NORMEIGONEDIMper{
\psi_{n}(\t)=\sum_{w\in S_N}\,(-1)^{l(w)}\,
\exp(2\pi i \sum_{j=1}^N (n_{w(j)}+\rho_{w(j)}) \t_j),
\qquad \,\,\,\,\,\,\,\,\,\,\,(n_1,\cdots,n_N) \in P_{++},}
\eqn\NORMILAIZATIONper{
\frac{1}{N!}\int_{(S^1)^N}\,\,d^N\t\,\,
 \overline{\psi_{n}}(\t)\,\psi_{n'}(\t)=\sum_{w\in S_N} \,(-1)^{l(w)}
\prod_{j=1}^N \delta_{n_{w(j)},n'_j}=\delta^{S_N}_{n,n'}.}
Here $\rho=(\rho_1,\cdots ,\rho_N)$ is a half-sum of the positive
roots of  $\frak{u}_N$.
 The integral kernel of the identity  operator can be represented as:
\eqn\KERNELper{
K(\t,\t')=K(\t-\t')=\delta_{(S_N)}(\t-\t')=
\sum_{n\in P_{++}}\,\, \overline{\psi_{n}}(\t)\,\psi_{n}(\t').}
The partition function of the Yang-Mills theory
on a torus $\Sigma_1$ is given by the trace of a evolution operator and:
\eqn\TRACEOP{
Z_{YM}(\Sigma_1)=
Tr\, e^{-t_2 H_2(\hat{p},\hat{q})}=\sum_{n\in
P_{++}}\,\,\int_{(S^1)^N}\,d^Nx \,\,
 \overline{\psi_{n}}(\t)\,e^{-t_2 H^{(0)}_2(i\pr_\t,\t)}\,\psi_{n}(\t).}
The kernel for the periodic case can be obviously represented as a matrix
element of the projection operator as follows:
\eqn\KERNELperiod{
K(\t,\t')=\int_{\IR^N/S_N}\, d^N\l\,\,
 \overline{\psi_{\l}}(\t)\,\,\,P(\l)\,\,\psi_{\l}(\t'),}
where the  wave-functions $\psi_{\lambda}(\t)$ are given by
\NORMEIGONEDIM\
and:
\eqn\PROJECTOR{
P(\l)=\sum_{m\in \IZ^N}\,\,\prod_{j=1}^N \delta(\l_j-m_j)=
\sum_{k\in \IZ^N}\,\,e^{2\pi i \sum_{j=1}^N\l_j k_j}.}
Equivalently we have:
\eqn\KERNELperiodsum{
K(\t,\t')=\sum_{k\in \IZ^N}
\int_{\IR^N/S_N}\, d^N\l\,\,
 \overline{\psi_{\l}}(\t)\,\,\,e^{2\pi i\sum_{j=1}^N \l_j
   k_j}\,\,\psi_{\l}(\t')=}
$$=\sum_{k\in \IZ^N}
\int_{\IR^N/S_N}\, d^N\l\,\,
 \overline{\psi_{\l}}(\t)\,\,\, \psi_{\l}(\t'+k).$$
We conclude that the Green function $G(\t,\t')$
(the path integral on the cylinder with insertion
of $\exp{(-t_2 H^{(0)}_2)}$)  is represented as:
\eqn\cyl{G_{YM}(\t,\t')= \sum_{n\in \IZ^N}\,\,\int_{\l\in
\IR^N/S_N}d^N\l\,\,
 \overline{\psi_{\l}}(\t)\, \,e^{2\pi i\sum_{j=1}^N \l_j n_j} e^{-t_2
p_2(\l)}\,\, \psi_{\l}(\t')}
or equivalently as:
\eqn\cylone{G_{YM}(\t,\t')= \sum_{k\in \IZ^N}\,\,\int_{\l\in
\IR^N/S_N}d^N\l\,\,
 \overline{\psi_{\l}}(\t) e^{-t_2 p_2(\l)}\,\,
\psi_{\l}(\t'+k).}
Let us  note that the identities  in \KERNELperiodsum, \cylone\ are
 based on the following transformation  property of the  complete
set of skew-symmetric normalized wave-functions on $\IR^N$:
\eqn\propert{\psi_{\lambda}(\t+k)=
\sum_{w\in S_N} (-1)^{l(w)}\,
e^{2\pi i \sum_{j=1}^N\l_{w(j)} k_j}
 e^{2\pi i \sum_{j=1}^N\l_{w(j)} \t_j}.}
Thus each elementary term in the averaging over $S_N$ is
multiplied on the simple exponent factor entering the description
of the projector \PROJECTOR. Let us also note that the shift
transformations in \propert\ can be interpreted as large gauge
transformations in Yang-Mills theory discussed above.

The representation \cylone\ can be written in the following form:
\eqn\green{G_{YM}(\t,\t')=\sum_{k\in \IZ^N}G_0(\t,\t'+k).}
If  we set the coupling $t_2$ to zero, $t_2=0$, we recover the
formula \KERNELperiodsum\ for
$K(\t,\t')$:
\eqn\for{K(\t,\t')=\sum_{k\in \IZ^N}K_0(\t,\t'+k).}
For the partition function of Yang-Mills theory on a torus we get
(after setting $\t=\t'$ above and integrating over $x$):
$$
Z_{YM}(\Sigma_1)=\sum_{n\in \IZ^N}\,\,\int_{\l\in \IR^N/S_N}d^N\l\,
 \int_{(S^1)^N}\,d^N\t \,\,
 \overline{\psi_{\l}}(\t)\, \,e^{2\pi i\sum_{j=1}^N \l_j n_j} e^{-t_2
p_2(\l)}\,\,
\psi_{\l}(\t)=
$$
$$
=\sum_{m\in P_{++}}\,\, e^{-t_2 p_2(m+\rho)}.
$$
and this coincides with the representation \YMMMsumone. Note that
obvious relations between \KERNELperiodsum,  \green, \for\  and the
averaging
over the topologically non-trivial gauge transformations discussed
above.

Now we are finally ready to consider the case of Yang-Mills-Higgs
theory. As it was conjectured above one can chose as a bases of the
wave-functions
the bases of eigenfunctions of the set of the Hamiltonian operators
in $N$-particle subsector of Nonlinear Schr\"{o}dinger theory.
Below we construct the Green function and partition function
in Nonlinear Schr\"{o}dinger theory
and demonstrate that identifying the Hamiltonian operator with
quadratic observable $\CO^{(0)}_2=\frac{1}{(2\pi)^2}{\rm Tr} \varphi_0^2$
in Yang-Mills-Higgs theory we reproduce the partition function of
Yang-Mills-Higgs theory on a torus.

Let us start with the construction of the kernel of the unit
operator in the bases of the $N$-particle eigenfunctions
of the Nonlinear Schr\"{o}dinger theory.
The representation for the kernel \KERNELperiodsum\
can be straightforwardly generalized to this case:
\eqn\KERNELperiodsum{
\widetilde{K}(x,x')=\sum_{(\l_1,\cdots \l_N)\in \CR_N}\
 \overline{\Phi_{\l}^{norm}}(x)
 \Phi^{norm}_{\l}(x')=\int_{\IR^N/S_N}\,\,d^N
\l\,\,\,\,\,\overline{\Phi_{\l}}(x)
\, P(\l) \,\Phi_{\l}(x'),}
 where $\Phi_{\l}(x)$ are normalized skew-invariant eigenfunctions on
$\IR^N$ given by \NORMWFUNC, $\Phi^{norm}_{\l}(x)$ are
normalized periodic eigenfunctions
 given by \PERIODNORM\ and the sum goes over the set $\CR_N$ of the
solutions of Baxter Ansatz equations.  The projector here is given by:
\eqn\PROJKERNELPERIOD{
P(\l)=\mu(\l) \,\sum_{m\in \IZ^N}\,\prod_{j=1}^N\,\,\delta(\a_j(\l)- m_j)=
\sum_{(\l^*_1,\cdots,\l^*_N)\in \CR_N}\prod_j\delta(\l_j-\l_j^*)}
where  $\a_j(\l)$ are defined as follows  (compare with \yang):
\eqn\YYANG{\alpha_j(\lambda)=\l_j+\frac{1}{2\pi i}
\sum_{k\neq j}\log \left(\frac{\l_k-\l_j-ic}{\l_k-\l_j+ic}\right).}
Then we have:
$$
\widetilde{K}(x,x')=\sum_{n\in \IZ^N} \int_{\IR^N/S_N}\,
d^N\l\,\,\mu(\l)\,\,
 \overline{\Phi}_{\l}(x)\,\,\,
e^{2\pi i\sum_{m=1}^N \l_m n_m}\,\prod_{l\neq j}
\left(\frac{\l_l-\l_j-ic}{\l_l-\l_j+ic}\right)^{n_j}
\,\,\Phi_{\l}(x')=$$
\eqn\PROJKERNELPERIODD{
=\sum_{k\in \IZ^N}
\int_{\IR^N/S_N}\, d^N\l\,\,
 \overline{\Phi}_{\l}(x)\,\,\, \Phi_{\l}(x'+n).}
The last equality is a consequence of  the following property
of the  eigenfunctions \NORMWFUNCAFF\ of the $N$-particle Hamiltonian in
Nonlinear Schr\"{o}dinger theory:
\eqn\NORMWFUNCAFFF{
\Phi_{\l}(x+n)=
\sum_{w\in W}(-1)^{l(w)}\,
\prod_{i<j}\left(\frac{\l_{w(i)}-\l_{w(j)}+ic}
{\l_{w(i)}-\l_{w(j)}-ic}\right)^{n_i}
\exp(2\pi i\sum_m\l_{w(m)}n_m)\times
}
$$
\times \prod_{i<j}\left(\frac{\l_{w(i)}-\l_{w(j)}+ic}
{\l_{w(i)}-\l_{w(j)}-ic}\right)^{\frac{1}{2}+[x_i-x_j]}
\exp(2\pi i\sum_k\l_{w(k)}x_k),$$
These wave functions are periodic and descend to the wave
functions \NORMWFUNC\
 if $\l=(\lambda_1,\cdots ,\lambda_N)$ satisfy \BAONE.

The representation for the kernel \PROJKERNELPERIODD\ leads to the
following
representation for the Green function (cylinder path integral) and
torus partition function for $t_{i\neq
2}=0$:
\eqn\cylo{G_{NLS}(x,x')=}
$$=\sum_{n \in \IZ^N} \int_{\IR^N/S_N}d^N\l\,\,\mu(\l)\,\,
\overline{\Phi_{\lambda}(x)}
\,e^{2\pi  i \sum_{m=1}^N\l_m n_m} \,\,
\prod_{l\neq j}
\left(\frac{\l_l-\l_j-ic}{\l_l-\l_j+ic}\right)^{n_j}e^{-t_2p_{2}(\l)}\Phi_{\lambda}(x'),
$$
or same:
$$G_{NLS}(x,x')=\sum_{k\in \IZ^N}G^0_{NLS}(x,x'+k)=\sum_{k\in \IZ^N}
\int_{\IR^N/S_N}\, d^N\l\,\,
 \overline{\Phi}_{\l}(x)e^{-t_2p_2(\l)} \Phi_{\l}(x'+k).$$
Similarly for kernel: $\widetilde K(x,x') =\sum_{k\in \IZ^N}
\widetilde K_0(x,x'+k)$  since the kernel is a Green function at $t_2=0$.
 Integrating over $x$ after setting $x=x'$
we obtain the representation for the partition function on the torus:
\eqn\tornew{Z_{NLS}(\Sigma_1)=}
$$=
\int_{\IR^N/S_N}d^N\l\,\,\mu(\l)\,\,\sum_{(n_1,\cdots ,n_N)\in \IZ^N}
e^{2\pi  i \sum_{m=1}^N\l_m n_m} \,\,
\prod_{l\neq j}
\left(\frac{\l_l-\l_j-ic}{\l_l-\l_j+ic}\right)^{n_j}e^{-t_2p_{2}(\l)}.$$
This is in a complete agreement with
a representation for the partition function
 of $U(N)$ Yang-Mills-Higgs theory  on a torus discussed in Section 3,
formula \SDHPART. Note that one can repeat the same arguments for
all observables and higher differential operators of Nonlinear
Schr\"{o}dinger theory,  traces of
higher powers of Dunkle operator from Section 4,
by simply turning on all other couplings $t_{k}$.
The identification of the representation of the partition function
of Nonlinear Schr\"{o}dinger operator and Yang-Mills-Higgs theory on
the torus  strongly suggests that the full equivalence of the
theories.

\subsec{More precise relation between wave-functions}

Now we  establish a precise relation between
wave-functions in Yang-Mills-Higgs $U(N)$ gauge theory and
$N$-particle sector in Nonlinear Schr\"{o}dinger equation theory.
Let us stress that the localization procedure in topological theories
is most straightforward in the case of compact manifolds.
Thus for example in the case of the torus in Yang-Mills-Higgs theory
we recover unambiguously the spectral invariants of the operators
$\CO_k^{(0)}$ acting on the Hilbert space of the theory.
On the classical level this means that the results do not depend on the
total derivative terms in the Lagrangian density.
It was shown above that the spectrum of operators $\CO_k^{(0)}=
\frac{1}{(2\pi i)^k}\Tr\,\varphi_0^k$
coincides with the spectrum of quantum Hamiltonians of Nonlinear
Schr\"{o}dinger theory restricted to $N$-particle sub-sector.
The identification of the bases of eigenfunctions implies a
fixing additional data - the choice of polarization for symplectic form
in the Hamiltonian quantization.
Below we
provide a choice of polarization in Yang-Mills-Higgs theory
leading to identification of wave-functions with eigenfunctions in
Nonlinear Schr\"{o}dinger theory. To simplify the presentation below we
omit symmetrization of wave-functions.

Let consider Yang-Mills-Higgs theory on $\Sigma_h$.
The result of abelianization
can be formulated in the form of an
effective gauge theory with the gauge group $U(1)^N$ and the
 bosonic part of the action  \SHITFS:
\eqn\SHITFSxx{\eqalign{ \Delta S=&\int_{\Sigma_h}\,d^2z
\,\sum_{i,j=1}^N \log\left(\frac{(\varphi_0)_i-(\varphi_0)_j+ic}{
    (\varphi_0)_i-(\varphi_0)_j-ic}
\right) F(A)^i +\cr
& +\frac{1}{2}\int_{\Sigma_h}\,d^2z
\,\sum_{i,j=1}^N \log((\varphi_0)_i-(\varphi_0)_j+ic)R^{(2)}\sqrt{g};}}
We consider a family of deformations \DEFACTNone\ of the theory:
\eqn\DEFACTNonexx{
\Delta S_{YMH}=-\sum_{k=1}^{\infty}\,t_k\,\,\int_{\Sigma_h}d^2z \,\,
\Tr\,\varphi_0^{k}\,\,{\rm
vol}_{\Sigma_h}=-\sum_{k=1}^{\infty}\sum_{j=1}^N t_k\int{(\varphi_0)}_j^k{\rm
vol}_{\Sigma_h}}
where we impose the condition that $t_k\neq 0$ for a finite
subset of the indexes $k\in\IZ$.

In the case of $\Sigma_h=T^2$ the Feynman path integral in
the  two-dimensional quantum field theory with the action \SHITFSxx\
is equivalent to the path integral in
 the one-dimensional theory  on $S^1$ with the phase space $T^*H$
supplied with the following symplectic form:
\eqn\omegadef{
\omega=d(\sum_{j=1}^N(\l_j+\sum_{k\neq j}\log\big(\frac{\l_j-\l_k+ic}
{\l_j-\l_k-ic}\big))d\t_j),\qquad \t_j\sim \t_j+2\pi i
n_j,\,\,\,\,n_j\in \IZ,}
which is a $c$-deformation
of the standard symplectic form $\omega=\sum_{j=1}^Nd\l_j\wedge d\t_j$
on $T^*H$ where $\{\t_j\}$ are  coordinates on $H$ and $\{\l_j\}$ are
coordinates in the fiber of the projection $T^*H\to H$.
Note that $\l_j$ can be identified with $(z,\az)$-independent $(\varphi_0)_i$. 

This symplectic form \omegadef\ can be easily transformed into a canonical one:
\eqn\omegadefcon{
\omega=d(\sum_{j=1}^N \a_j(\l)d\t_j)=\sum_{j=1}^N d\a_j(\l)\wedge
d\t_j,}
where
\eqn\idef{\a_j(\l)=\l_j+\sum_{k=1,k\neq j}^N\log\big(\frac{\l_j-\l_k+ic}
{\l_j-\l_k-ic}\big).}
Quantization in $\t$-polarization is given by the following
realization:
\eqn\polq{
\widehat{\a}_j=-i\frac{\pr}{\pr \t_j},\qquad \widehat{\t}_j=\t_j.}
A bases of wave functions then
can be constructed using a complete set of common eigenfunctions of
the operators $\widehat{\a}_j$:
\eqn\wavefunc{
\Psi_{\l}(\t)=e^{i\sum_{j=1}^N \a_j(\l)\t_j}=\prod_{j,k=1}^N
\big(\frac{\l_j-\l_k+ic}{\l_j-\l_k-ic}\big)^{\t_j}\,e^{i\sum_{j=1}^N\l_j\t_j},}
\eqn\wvaefunceig{\widehat{\a}_j\,\Psi_{\l}(\t)=\a_j(\l)\,\Psi_{\l}(\t).}
The deformation \DEFACTNonexx\ leads to a consideration
of  the following set of Hamiltonians in the effective
one-dimensional theory:
\eqn\Hamsim{\CH_k=\sum_{j=1}^N\widehat{\l}_j^k,}
where $\hat \lambda$ is related to the derivative in $\theta$ through:
\eqn\eigendef{\widehat{\a}_j=
\widehat{\l}_j+\sum_{k\neq j}\log\Big(\frac{\widehat{\l}_j
-\widehat{\l}_k+ic} {\widehat{\l}_j-\widehat{\l}_k-ic}\Big)=-i{\partial
\over {\partial \theta_j}}.}
Computations from Section 3 show that the eigenvalues of the Hamiltonians
are given by $E_k=\sum_{j=1}^N\lambda_j^k$ with $\lambda$'s subject to
(for periodicity
conditions $\theta_j\sim \theta_j+2\pi m_j$, $m_j\in \IZ$):
\eqn\eigendefex{\a_j(\l)=
\l_j+\sum_{k\neq j}\log\Big(\frac{\l_j
-\l_k+ic} {\l_j-\l_k-ic}\Big)=n_j, \qquad n_j\in \IZ.}
Let us remark that in $\t$-representation  Hamiltonians \Hamsim\
act on wave-functions  by complicated integral operators
due to the non-trivial relations \idef\ between $\widehat{\a}_j$ and
$\widehat{\l}_k$.

Compare now the bases \wavefunc\ of common  eigenfunctions of $\CH_k$,
$k\in \IZ_+$  with the bases of common eigenfunctions for the Yang
$N$-particle system:
\eqn\wavefuncY{
\Psi^Y_{\l}(x)=\prod_{j,k=1}^N\Big(\frac{\l_j-\l_k+ic}
{\l_j-\l_k-ic}\Big)^{[x_j]}\,e^{i\sum_{j=1}^N\l_jx_j},}
where  $[x]$ is an integer part of $x$. Here we also 
impose  periodicity conditions
$x_j\sim x_j+2\pi m_j$, $m_j\in \IZ$ and thus: 
\eqn\BAquant{ \l_j+\sum_{k\neq  j}\log\Big(\frac{\l_j-\l_k+ic}
{\l_j-\l_k-ic}\Big)=n_j,\qquad n_j\in \IZ.}

The Yang system can be considered as a quantization of $T^*H$ with the
symplectic structure:
\eqn\symYanf{
\omega=\sum_{j=1}^N dp_j\wedge dx_j,\qquad x_j\sim x_j+2\pi n_j,\quad
n_j\in
\IZ.}
In $x$-polarization we have:
\eqn\polx{ 
\widehat{p}_j=-i\frac{\pr}{\pr x_j},\qquad \widehat{x}_j=x_j.}
The Hamiltonians $\CH_k$ of Yang system  can be
expressed as symmetric functions of Dunkle operators $\CD_j$
(see Section 4), and for $k=2$ we have:
\eqn\CHtwo{
\CH_2=\frac{1}{2}\sum_{i=1}^N\CD_j^2=
\frac{1}{2}\sum_{i=1}^N\widehat{p}_j^2
+c\sum_{1\leq i<j\leq N}\delta(x_i-x_j),}
The eigenvalues $E_k$ of the Hamiltonians $H_k$ are given by:
\eqn\HamY{E_k=\sum_{j=1}^N\l_j^k,\qquad
  \l_j+\sum_{k\neq j}\log\Big(\frac{\l_j-\l_k+ic}
{\l_j-\l_k-ic}\Big)=n_j,}
and  coincides with the spectrum in Yang-Mills-Higgs theory.

The two bases of wave functions \wavefunc\ and
\wavefuncY\ are related by a unitary operator with the 
integral kernel given by: 
\eqn\kernel{
K(\t,x)=\sum_{{\bf n}\in \IZ^N} \Psi_{\l_{{\bf n}}}(\t)
\overline{\Psi^Y_{\l_{{\bf n}}}}(x),}
where the sum is over solutions $\l_{{\bf n}}=(\l_1({\bf
  n}),\ldots,\l_N({\bf n}))$  of the equations \BAquant\
for all ${\bf n}=(n_1,\ldots,n_N)\in \IZ^N$. The only invariant that
does not depend on the choice of the bases of wave functions is the
common spectrum of the operators $\CH_k$ which is captured  by 
partition function. The coincidence of the partition functions for 
Yang-Mills-Higgs theory and Yang system was demonstrated in the
previous subsection.    

Let us stress that in $\t$-representation  there is no simple way to write
eigenfunction problem for wave-functions in terms of the differential
operators while in $x$-representation the eigenfunction problem for
Hamiltonians $\CH_k$ acting on $\Psi^Y_{\l}(x)$ is
given by  the Yang equations and its higher-derivative analogs.
This explains the relevance of the $x$-polarization in
Yang-Mills-Higgs theory deformed by local operators $\CO^{(0)}_k$.

\newsec{Generalization  of $G/G$ gauged WZW model}

In this section we describe following \GS\
another instance of the relation
between two-dimensional gauge theories and
quantum integrability in many-body integrable systems.
We consider a generalization of GWZW theory
and demonstrate that  partition function can be represented as a sum
over solutions of a certain generalization of Bethe Ansatz equation.

We start with the definition of the
set of fields and the action of the odd and even symmetries
in the spirit of \UONEACT, \GAUGEYM, \BRST. Let us note that
the gauged Wess-Zumino-Witten  model can be obtained
from the topological Yang-Mills theory by using the group-valued field $g$
instead of algebra-valued field $\varphi$. Introduce the
set of fields $(A,\psi_A,\Phi,\psi_{\Phi},\chi_{\pm},\varphi_{\pm},g)$
and $t\in \IR^*$ with the following action of the
odd and even symmetries:
\eqn\EVENGGWZW{
\CL_{(g,t)}\,A^{(1,0)}=(A^g)^{(1,0)}-A^{(1,0)}
\,\,\,\,\,\,\,\CL_{(g,t)}\,A^{(0,1)}_A=-(A^{g^{-1}})^{(0,1)}+A^{(0,1)},}
$$\CL_{(g,t)}\,\psi_A^{(1,0)}=-g\psi_A^{(1,0)}
g^{-1}+\psi_A^{(1,0)},\,\,\,\,\,\,
\CL_{(g,t)}\,\psi_A^{(0,1)}=g^{-1}\psi_A^{(0,1)}
g-\psi_A^{(0,1)},\,\,\,\,\CL_{(g,t)}\,g=0,$$
$$\CL_{(g,t)}\,\Phi^{(1,0)}=tg\Phi^{(1,0)}
g^{-1}-\Phi^{(1,0)},\,\,\,\,\,\,
\CL_{(g,t)}\,\Phi^{(0,1)}=-t^{-1}g^{-1}\Phi^{(0,1)} g+\Phi^{(0,1)},
$$
$$
\CL_{(g,t)}\,\psi_{\Phi}^{(1,0)}=tg\psi_{\Phi}^{(1,0)}
g^{-1}-\psi_{\Phi}^{(1,0)},
\,\,\,\,\,\,
\CL_{(g,t)}\,\psi_{\Phi}^{(0,1)}=-t^{-1}g^{-1}\psi_{\Phi}^{(0,1)}
g+\psi_{\Phi}^{(0,1)},
$$
$$\CL_{(g,t)}\,\chi_{+}=tg\chi_+g^{-1}-\chi_+,\,\,\,\,\,\,
\CL_{(g,t)}\chi_-=-t^{-1}g^{-1}\chi_-g+\chi_- $$
$$
\CL_{(g,t)}\,\varphi_{+}=t^{-1}g\varphi_{+}g^{-1}-\varphi_{+},\,\,\,\,\,\,
\CL_{(g,t)}\varphi_{+}=-tg^{-1}\varphi_{+}g+\varphi_{+}
$$
\eqn\ODDGGWZW{
Q\,A=i\psi_A,\,\,\,\,\,\,Q\,\psi^{(1,0)}_A=i(A^g)^{(1,0)}-iA^{(1,0)},
\,\,\,\,\,\,\,\,Q\,\psi^{(0,1)}_A=-i(A^{g^{-1}})^{(0,1)}+iA^{(0,1)},}
$$Q\,g=0,$$
\eqn\BRSTGGWZW{
Q\Phi=i\psi_{\Phi},\,\,\,\,
Q\psi^{(1,0)}_{\Phi}=t g\Phi^{(1,0)} g^{-1}-\Phi^{(1,0)},
\,\,\,\,\,Q\psi^{(0,1)}_{\Phi}=-t^{-1}g^{-1}\Phi^{(0,1)} g-\Phi^{(0,1)},}
$$Q\chi_{\pm}=i\varphi_{\pm},\,\,\,\,
Q\varphi_+=t g\chi_+ g^{-1}- \chi_+,\,\,\,\,\,\,
Q\varphi_-=-t^{-1} g^{-1}\chi_- g+ \chi_- .$$
We have $Q^2=\CL_{(g,t)}$ and $Q$ can be considered as a
BRST operator on the space of $\CL_{(g,t)}$-invariant
functionals.

We define the action of the theory in analogy with the construction
of the action for Yang-Mills-Higgs theory as follows:
\eqn\ACTONQCOM{
S=S_{GWZW}+[Q,\int_{\Sigma_h}\,d^2z \,\Tr\,( \frac{1}{2}
\Phi\wedge  \psi_{\Phi}+}
$$+\tau_1\,(\varphi_+
 \nabla^{(1,0)}_A\Phi^{(0,1)}+\varphi_-
 \nabla^{(0,1)}_A\Phi^{(1,0)})+\tau_2(\chi_+\varphi_-+\chi_-\varphi_+)
 {\rm vol}_{\Sigma_h})]_+.$$
Taking $\tau_1=0$, $\tau_2=1$ and
 applying  the  standard localization technique to this theory
 we obtain for the partition function:
\eqn\GWZWHPART{
Z_{GWZWH}(\Sigma_h)=\frac{e^{(1-h)a(t)}}{|W|}
\int_{H} d^N\l\,\,\mu_q(\l)^{h}\,\sum_{(n_1,\cdots,n_N)\in \IZ^N}
e^{2\pi i \sum_{m=1}^N\l_m n_m(k+c_v)}\times }
$$\times  \prod_{j\neq k}(e^{2\pi i(\l_j-\l_k)}-1)^{n_j-n_k+1-h}
\,\,\prod_{j,k}(te^{2\pi i(\l_j-\l_k)}-1)^{n_j-n_k+1-h},$$
where $a(t)$ is a $h$-independent constant, the integral
goes over the Cartan  torus $H=(S^1)^N$ and
\eqn\QMEASURE{
\mu_q(\l)=\det\|\frac{\pr \beta_j(\l)}{\pr \l_k}\|
,}
with:
\eqn\newone{e^{2\pi i \beta_j(\lambda)}=e^{2\pi i
\l_j(k+c_v)}\,\prod_{k\neq
  j}\frac{te^{2\pi i (\l_j-\l_k)}-1}{te^{2\pi i (\l_k-\l_j)}-1}.}
We can rewrite this formula in the form similar to
\SDHsumon:
\eqn\GWZWHPARTo{
Z_{GWZWH}(\Sigma_h)=\frac{e^{(1-h)a(t)}}{|W|}
\int_{H} d^N\l\,\,\mu_q(\l)^{h}\,\sum_{(n_1,\cdots,n_N)\in \IZ^N}
e^{2\pi i \sum_{m=1}^N \beta_m(\lambda) n_m}\times }
$$\times  \prod_{j< k}(e^{i\pi(\l_j-\l_k)}-e^{i\pi(\l_k-\l_j)})^{2-2h}
\,\,\prod_{j<k}|te^{i\pi (\l_j-\l_k)}-e^{i\pi(\l_k-\l_j)}|^{2-2h}.$$
Summation over
integers in \GWZWHPART\ leads to the following restriction on the
integration parameters:
\eqn\QBA{
e^{2\pi i \l_j(k+c_v)}\,\prod_{k\neq
  j}\frac{te^{2\pi i (\l_j-\l_k)}-1}{te^{2\pi i (\l_k-\l_j)}-1}=1
\,\,\,\,,\,\,\,\,\,\,\, i=1,\cdots ,N,}
It is useful to rewrite the equations \QBA\ in the standard form
of the Bethe Ansatz equations:
\eqn\QBAsin{
e^{2\pi i \l_j(k+c_v)}\,\prod_{k\neq
  j}\frac{sin(i\pi (\l_j-\l_k +ic))}{sin(i\pi (\l_j-\l_k-ic))}=1
\,\,\,\,,\,\,\,\,\,\,\, i=1,\cdots ,N,}
This clearly shows that we are dealing with a kind of XXZ quantum
integrable chain. The particular form \QBAsin\ can be obtained
by the taking the limit $s\to -i\infty$ in the following Bethe
equations:
\eqn\QBAsin{
\left(\frac{sin(i\pi (\l_j-i s c ))}{sin(i\pi (\l_j+i s c))}\right)
^{(k+c_v)}\,\prod_{k\neq
  j}\frac{sin(i\pi (\l_j-\l_k +ic))}{sin(i\pi (\l_j-\l_k-ic))}=1
\,\,\,\,,\,\,\,\,\,\,\, i=1,\cdots ,N,}
corresponding to formal limit of the infinite spin $s$ of XXZ chain.

The partition function is the generalization of Yang-Mills-Higgs
theory, discussed above, and can be written in the following form:
\eqn\PARTDISCRQ{
Z_{GWZWH}(\Sigma_h)=\sum_{\l_i\in \CR_q}\,\,(D_{\l}^q)^{2-2h},}
where $\CR_q$ is a set of the solutions of \QBA\ and:
\eqn\SDHDIMQ{
D^q_{\l}=\mu_q(\l)^{-1/2}
\prod_{i<j}(q^{\frac{1}{2}(\l_i-\l_j)}-q^{\frac{1}{2}(\l_j-\l_i)})
\prod_{i<j}|t q^{\frac{1}{2}(\l_i-\l_j)}-q^{\frac{1}{2}(\l_j-\l_i)}|,}
where  we use the standard parametrization $q=\exp(2\pi i/(k+c_v))$.
Note that in the limit $t\to \infty$ equation \QBA\
and the expression for the partition function \SDHDIMQ\
up to an  overall scaling factor become the corresponding expressions
for a  gauged Wess-Zumino-Witten model. Finally note that the form of
\QBA\
and the explicit expressions for the q-Casimir operators, playing the
role of the Hamiltonians, strongly imply the description of the wave
functions of the theory in terms of the wave functions in a particular
$XXZ$ finite spin chain. This proposition will be discussed in details
elsewhere.

{\bf  Acknowledgements}.  We are grateful to J. Bernstein, M.
Kontsevich, W. Nahm, N. Nekrasov, F. Smirnov and L. Takhtajan for
discussions. The research is  supported by  Science Foundation Ireland
grant.

\listrefs
\end